\documentclass[12pt]{article}
\usepackage{amsmath,amssymb,amscd,cite}
\newtheorem{thm}{Theorem}[section]
\newtheorem{prop}[thm]{Proposition}
\newtheorem{lem}[thm]{Lemma}
\newtheorem{cor}[thm]{Corollary}

\newcommand{\pf}{{\bf Proof. \ }}
\newcommand{\qed}{\hfill $\blacksquare$ \\}

\font\msbm=msbm10 at 12pt

\newcommand{\F}{\mbox{\msbm F}}

\newtheorem{ex}[thm]{Example}

\newcommand{\ord}{ord}
\date{}

\begin{document}

\title{On Repeated-Root Constacyclic Codes of Length $2^amp^r$ over Finite Fields.}

\author{Aicha Batoul, Kenza Guenda and T. Aaron Gulliver\thanks{A. Batoul and K. Guenda
are with the Faculty of Mathematics USTHB, University of Science and
Technology of Algiers, Algeria. T. Aaron Gulliver is with the
Department of Electrical and Computer Engineering, University of
Victoria, PO Box 3055, STN CSC, Victoria, BC, Canada V8W 3P6.
email: a.batoul@hotmail.fr, kguenda@gmail.com, agullive@ece.uvic.ca.}}
\maketitle
\begin{abstract}
In this paper we investigate the structure of repeated root constacyclic
codes of length $2^amp^r$ over $\mathbb{F}_{p^s}$ with $a\geq1$ and $(m,p)=1$.
 We characterize the codes in terms of their
generator polynomials. This provides simple conditions on the
existence of self-dual negacyclic codes. Further, we gave cases where the constacyclic codes are equivalent to cyclic codes.

\end{abstract}


\section{Introduction}
Constacyclic codes over finite fields form a remarkable class of linear codes, as they include the important family of cyclic
codes. Constacyclic codes also have practical applications as they can be efficiently encoded using simple shift registers. They
have rich algebraic structures for efficient error detection and correction. This explains their preferred role in engineering.
Repeated-root constacyclic codes, were first studied in 1967 by Berman \cite{Berman}, then
by several authors such as Falkner et al \cite{Falkner} and Salagean \cite{Salagean}.
Repeated-root cyclic codes were first investigated in the most generality in the 1990s by Castagnoli et al \cite{Castagnoli},
and van Lint \cite{vanlint}, where they showed that repeated-root cyclic codes have a concatenated construction,
and are not asymptotically good. However, it turns out that optimal repeated-root constacyclic codes
still exist. 
These motivate researchers to further study this class of codes.

Recently, Dinh, in a series of papers \cite{dinh1},\cite{dinh2} and \cite{dinh3}, determined the generator polynomials of all constacyclic codes over $\F_q$, of lengths $2p^r$, $3p^r$ and $6p^r$. These results have been extended to more general code lengths. The generator polynomials
of all constacyclic codes of length $2^tp^r$ over $\F_{p^s}$ were given in \cite{Bakshi}.
The generator polynomials of all constacyclic codes of length $lp^r$ over $\F_q$ were characterized in \cite{Chen}, where $l$ is a prime different from $p$.

In this paper, we extend the main results of Batoul et al given in \cite{BGGC} and of Guenda and Gulliver given in \cite{GG} to constacyclic codes of length $2^amp^r$ over $\F_{p^s}$ where $a\geq1$ and $m$ is an odd integer with $(m,p)=1$.
The remainder of the paper is organized as follows. Some preliminary
results are given in Section 2. In Section 3, the structure of generator of constacyclic
codes of length $mp^r$ is given using the generator polynomial of
constacyclic codes of length $m$. In Section 4, the structure of generator of constacyclic
codes of length $2^amp^r$ is given. Further, we gave cases where the constacyclic codes are equivalent to cyclic codes.  It is well know that the only self-dual constacyclic codes over finite fields are either cyclic codes over fields with even characteristic  or negacyclic codes.
For that in Section 5, we give conditions on the
existence of self-dual negacyclic codes of length $2^amp^r$ over $\F_{p^s}$, where $p$ is odd. 

\section{Preliminaries}
Let $p$ be a prime number and $\mathbb{F}_{q}$ the finite field
with $q=p^s$ elements. We note its group of units  $\F_q^*$.

The order of an element $a$ in the multiplicative group $\F_q^*$ is the least integer $b$ such that
$a^b=1$ in $\F_q^*$, we note $b$ by $\ord_q (a)$.
Let $i$ be an integer with $0\leq i <n$, the $q$-cyclotomic coset of $i$ modulo $n$ is the set
\[ C_i=\{i,iq,\ldots,iq^{l-1}\}\,\,(\mod\,n),\]
where $l$ is the smallest positive integer such that $iq^l\equiv\,i(\mod n)$.

The minimal polynomial of $\beta^i$ over $\F_q$ is
\[M_{\beta^i}(x)=\prod_{j\in C_i}(x-\beta^j),\]
where $C_i$ is the $q$-cyclotomic coset modulo $n$ and $\beta$ is a primitive element of $\F_q$.

 An $[n,k]$ linear code $C$ over $\mathbb{F}_{p^s}$ is a $k$-dimensional subspace of
$\mathbb{F}_{p^s}^n$.
For $\lambda$ in $\F_q^*$, a linear code $C$ of length $n$ over $\F_q$ is said to be constacyclic if it satisfies
\[
(\lambda c_{n-1},c_{0},\ldots,c_{n-2})\in C,\text{ whenever }(c_{0},c_{1},
\ldots, c_{n-1}) \in C.
\]

 When $\lambda =1$ the code is called cyclic, and when $\lambda=-1$ the code is called negacyclic.
The Euclidean dual code $C^{\bot}$ of $C$ is defined as
$C^{\bot}=\{\mathsf{x} \in \mathbb{F}_q^n;\,\, \sum_{i=1}^{n}x_iy_i=0
\, \forall \, \mathsf{y} \in C\}$. An interesting class of codes is
the so-called self-dual codes. A code is called Euclidean self-dual
if it satisfies $C=C^{\bot}$. Note that the dual of a $\lambda$-constacyclic code
is a $\lambda^{-1}$-constacyclic code.
A monomial linear transformation of $\F_{q}^n$ is an $\F_q$-linear
 transformation $\tau$ such that there exists scalars $\lambda_1,
\ldots,\lambda_n$ in $\F_{q}^{*}$ and a permutation $\sigma \in S_n$
(the group of permutation of the set $\{1,2,\ldots,n\}$) such that,
 for all $(x_1,x_2,\ldots,x_n)\in \F_{q}^n$, we have
\[\tau(x_1,\ldots,x_n)= (\lambda_1 x_{\sigma (1)}, \lambda_2
x_{\sigma (2)}, \ldots, \lambda_n x_{\sigma (n)}).\]
Two linear codes $C$ and $C'$ of length $n$ are called monomially equivalent if there
exists a monomial transformation of $\F_{q}^n$, such that
$\tau(C)=C'$.
Here, whenever two codes are said to be equivalent it is meant that they are monomially equivalent.
Let $C$ be a $\lambda$-constacyclic,then $C$ is an ideal of the quotient ring
$R_n=\mathbb{F}_{p^s}[x]/\langle x^n-\lambda \rangle$.
It is well known that every $\lambda$-constacyclic code is generated by a unique polynomial
of least degree. Such a polynomial is called the generator of the code and it is a divisor of $x^n-\lambda$.
Therefore, there is a one-to-one correspondence between constacyclic codes of length $n$ over $\F_q$,
and divisors of $x^n-\lambda$.\\
\section{Constacyclic Codes of Length $mp^r$ over $\mathbb{F}_{p^s}$}
Throughout this section $p$ is an odd prime number and $n=mp^r$, with $m$ an integer
such that $(m,p)=1$.

This section provides the structure of constacyclic codes of length $mp^r$
over $\F_q$. We give this important lemma which we need after.
\begin{lem}
\label{lem:puissance} Let $q=p^s$ be a prime power. Then for all $
\lambda$ in $\F_q^{*}$ there exist  $\lambda_0$ in $\F_q^{*}$, such that $\lambda=\lambda_0^{p^r}$, for $r$ in $\mathbb{N}$.
\end{lem}
\pf Let $\lambda \in \F_{q}^{*}$, $q=p^s$ and $s$ be a positive integer.
 If $s<r$ then there exists integers $k$ and $m$ such that $r-m =ks$ with $0\leq m \leq s-1,$  $s-m= s-(r-ks)=s(k+1)-r$.
Let $\lambda_0=\lambda^{p^{s-m}},$ then  $\lambda_0^{p^r}=(\lambda^{p^{s(k+1)-r}})^{p^r}=(\lambda^{p^{s(k+1)}})=\lambda.$
 If $r<s$, then $(\lambda^{p^{s-r}})^{p^{r}}= \lambda$.
\qed

 It is well know that $\lambda$-constacyclic codes over $\F_q$ are principal ideals
 generated by factors of $x^{mp^r}-\lambda$. Since $\F_{p^s}$ has characteristic $p$,
 and by Lemma~\ref{lem:puissance} the polynomial $x^{mp^r}-\lambda$ can be factored as
 \begin{equation}
\label{eq:equation1} x^{mp^r}-\lambda=x^{mp^r}-\lambda_0^{p^r}=(x^{m}-\lambda_0)^{p^r}.
\end{equation}
 The polynomial $x^{m}-\lambda_0$ is a monic square free polynomial. Hence
from~\cite[Proposition 2.7]{permounth} it factors uniquely as a
product of pairwise coprime monic irreducible polynomials $f_0(x),\ldots, f_l(x)$. Thus from~(\ref{eq:equation1}) we obtain the
following factorization of $x^{mp^r}-\lambda_0$.
\begin{equation}
\label{eq:equation2} x^{mp^r}-\lambda_0^{p^r}={f_0(x)}^{p^r}\ldots {f_l(x)}^{p^r}.
\end{equation}
A $\lambda$-constacyclic  code of length $n=mp^r$ over $\mathbb{F}_{p^s}$ is then
generated by a polynomial of the form
\begin{equation}
\label{eq:gen}
 A(x)=\prod {f_i}^{k_i},
 \end{equation}
where $f_i(x), 0\leq i \le l$, are the polynomials given in
(\ref{eq:equation2}) and $0 \le k_i \le p^r$.

\section{Constacyclic Codes of Length $2^amp^r$ over $\mathbb{F}_{p^s}$}
\label{section:3}
In this section we first recall the following important result of Batoul et al. given in \cite{BGGC}
\begin{prop}\cite[Proposition 3.2]{BGGC}
\label{propostion3.2}
 Let $q$ be a prime power, $n$ a positive integer and
$\lambda$ an element in $\F_q^*$. If $\F_{q}^{*}$ contains an element $\delta$, where $\delta$ is an $n$-th root of
$\lambda$, then a $\lambda$-constacyclic code of length $n$ is equivalent to a cyclic code of length $n$.
\end{prop}
 And we give the structure of repeated-root constacyclic
codes over $\F_q$,$q=p^s$ of length $2^amp^r$, $a\geq1$.
But before that and in the goal of using the isomorphism between
cyclic codes and constacyclic codes of the same length, given in Proposition \ref{propostion3.2},
we give the structure of cyclic codes of length $2^amp^r$ over $\F_q$,
 for that, we need the following Lemma:
\begin{lem}
\label{lem:primitive}
Let $a\geq1$ and  $\alpha$  a primitive $2^a$-th root  of the unity in $\F_q^*$,the following holds:
\begin{enumerate}
\item [1)] $\alpha^{2^i}$ is a primitive $2^{a-i}$-th root of the unity in $\F_q^*$ for all $i,\,\,i\leq a$.
  \item [2)] $\alpha^{m}$ is a primitive $2^a$-th root of the unity in $\F_q^*$ for all odd integer $m$.
  \item [3)] $\prod_{k=1}^{2^a}\alpha^k=-1$.
\end{enumerate}
\end{lem}
\pf
\begin{enumerate}
  \item [1)]Let $i,\,\,i\leq a$, in the cyclic group $\F_q^*$, we have that $\ord(\alpha^{2^i})=
\frac{\ord(\alpha)}{(2^i,\ord(\alpha))}=\frac{2^a}{(2^i,2^a)}=\frac{2^a}{2^i}=2^{a-i}$.\\
  \item [2)]Since $(2^a,m)=1$, so  $\ord(\alpha^m)=
\frac{\ord(\alpha)}{(m,\ord(\alpha))}=\frac{2^a}{(m,2^a)}=2^a$.\\
  \item [3)] $(x^{2^a}-1)=\prod_{k=1}^{2^a}(x-\alpha^k)$ then $\prod_{k=1}^{2^a}\alpha^k=(-1)^{2^{a}}\frac{(-1)}{1}=-1$.
\end{enumerate}
\qed

\begin{prop}
\label{prop:decoposition}
 Let $q$ be a power of an odd prime $p$ and $n=2^am$ a positive
integer such that $m$ is an odd integer and $(m,p)=1$,$a\geq1$.
 Then if $\F_{q}^{*}$ contains a primitive
$2^a$-root of unity $\alpha$ and the $f_i(x)$, $0 \le i \le l$ are
the monic irreducible factors of $x^{m}-1 $ in $\F_{q}[x]$,then:
\begin{equation}
\label{eq:decoposition}
x^{2^am}-1= \prod_{k=1}^{2^a}(\prod_{i=0}^{l}f_{i}(\alpha^{-k}x)).
\end{equation}
\end{prop}
\pf Assume that $x^{m}-1 =\prod_{i=0}^{l}f_{i}(x)$ is the
factorization of $x^{m}-1$ into monic factors over $\F_{q}$.
This factorization is unique since it is over a unique factorization
domain (UFD).\\
Let $\alpha \in\F_q^*$ be a primitive $2^a$-th root of unity and let $1\leq k\leq2^a$.
\[
\begin{array}{ccl}
(\alpha^{-k}x)^m-1&=& \prod_{i=0}^{l}f_{i}(\alpha ^{-k}x)\\
(\alpha^{-k})^m(x^m-(\alpha^{k})^m)& =&\alpha^{-k}\prod_{i=0}^{l}f_{i}(\alpha ^{-k}x)\\
(x^m-\alpha^{km})&=& \alpha^{k(m-1)}\prod_{i=0}^{l}f_{i}(\alpha ^{-k}x)\\
(x^m-(\alpha^{m})^k)&=& \alpha^{k(m-1)}\prod_{i=0}^{l}f_{i}(\alpha ^{-k}x)
\end{array}
\]
Then by Lemma~\ref{lem:primitive} $\alpha^{m}$ is also
a primitive $2^a$-th root of unity, we obtain:
\[
\begin{array}{ccl}
\prod_{k=1}^{2^a}(x^m-(\alpha^{m})^k)&=&\prod_{k=1}^{2^a}\alpha^{k(m-1)}\prod_{i=0}^{l}f_{i}(\alpha ^{-k}x)\\
&=&\prod_{k=1}^{2^a}\alpha^{k(m-1)}\prod_{k=1}^{2^a}\prod_{i=0}^{l}f_{i}(\alpha ^{-k}x)\\
& =&\left( \prod_{k=1}^{2^a}\alpha^{k(m-1)}\right)  \left( \prod_{k=1}^{2^a}\prod_{i=0}^{l}f_i(\alpha^{-k}x)\right)  \\
& =&\left( \prod_{k=1}^{2^a}\dfrac{\alpha^{km}}{\alpha^k}\right) \left( \prod_{k=1}^{2^a}\prod_{i=0}^{l}f_i(\alpha^{-k}x)\right)
\end{array}
\]
Since $(x^{2^am}-1)=((x^m)^{2^a}-(\alpha^m)^{2^a})=\prod_{k=1}^{2^a}(x^m-\alpha^{km})$ we obtain the result:
\[x^{2^am}-1= (\prod_{k=1}^{2^a}(\prod_{i=0}^{l}f_{i}(\alpha^{-k}x)).\]
\qed
\begin{cor}
\label{cor:prod} Let $q$ be a power of an odd prime $p$ and $n=2^amp^r$ a positive
integer such that $m$ is an odd integer and $(m,p)=1$, $a\geq1$.
 Then if $\F_{q}^{*}$ contains a primitive
$2^a$-root of unity $\alpha$ and the $f_i(x)$, $0 \le i \le l$ are
the monic irreducible factors of $x^{m}-1 $ in $\F_{q}$ then:
\[(x^{2^amp^r}-1)=(x^{2^am}-1)^{p^r}=\prod_{k=1}^{2^a}\prod_{i=0}^{l}f_{i}^{p^r}(\alpha ^{-k}x).\]
\end{cor}
\pf Since the characteristic of $\F_q$ is $p$, the proof follows from Proposition \ref{prop:decoposition}.
\qed

In the following we give the structure of cyclic codes of length $2^amp^r$ over $\F_q$
\begin{cor}
\label{cor:structure cyclic}
Let $q$ be a power of an odd prime $p$, $n=2^amp^r$ be a positive
integer such that $m$ is odd integer, with $a\geq1$ and $(m,p)=1$.
 Then if $\F_{q}^{*}$ contains a primitive
$2^a$-root of unity $\alpha$ and the $f_i(x)$, $0\le i \le l$ are
the monic irreducible factors of $x^{m}-1 $ in $\F_{q}[x]$ then any cyclic code
of length $n=2^amp^r$ is generated by $\prod_{k=1}^{2^a}(\prod_{i=0}^{l}f_{i}^{j_i}(\alpha ^{-k}x))$
where $0\leq j_i\leq p^r$.
\end{cor}
\pf
Since any cyclic code of length $n=2^amp^r$ is generated by a divisor of $(x^{2^amp^r}-1)$, hence
by Corollary \ref{cor:prod} we have that
\[(x^{2^amp^r}-1)=(x^{2^am}-1)^{p^r}=\prod_{k=1}^{2^a}\prod_{i=0}^{l}f_{i}^{p^r}(\alpha ^{-k}x)\]
So we deduce the result.
\qed

Now we generalize Proposition \ref{propostion3.2}.

\begin{thm}
\label{co:equiv4} Let $q$ be a power of an odd prime $p$
 and $m$ an odd integer such that $(m,p)=1$. Let $\lambda$ and $\delta$ in the multiplicative group
$\F_{q}^{*}$ such that $\delta^{m}=\lambda$, if $\delta =\beta^{2^a}$ in $\F^*_{q}$, then the following hold:
\begin{enumerate}
\item [(i)] The $ \lambda$-constacyclic codes of length $2^am p^{r}$ over $\F_q$
are equivalent to cyclic codes of length $2^am p^{r}$ over $\F_q$.
\item [(ii)] If $q \equiv 1 \mod 2^{a+1}$, then $-\lambda$-constacyclic codes of length $2^am p^{r}$ over $\F_q$
are equivalent to cyclic codes of length $2^am p^{r}$ over $\F_q$.
\end{enumerate}
\end{thm}
\pf
 For the part (i), let $\lambda \in\F_{q}^{*}$ such that there exists $\delta \in \F_{q}^{*}$,
$\delta^{m}=\lambda$ and $\delta =\beta^{2^a}$ in $\F_{q}$ then
$\lambda =\beta^{2^am}$. So by Lemma \ref{lem:puissance} there exists $\beta_0\in \F_{q}^{*}$ such that $\beta=\beta_0^{p^r}$.
Then $\lambda =\beta_0^{2^amp^r}$, hence  by Proposition \ref{propostion3.2}, $\lambda$-constacyclic codes of length $2^am p^{r}$ over $\F_q$
are equivalent to cyclic codes over $\F_q$.
For the part (ii), since $q \equiv 1 \mod 2^{a+1}$ and by Lemma~\ref{lem:primitive} there exists a primitive $2^{a+1}$-root of unity
 $\alpha \in\F_q^*$. So $\alpha^{2^a}=-1$ and $-\lambda=(-1)^{mp^r}\lambda=(\alpha^{2^a})^{mp^r} \beta_0^{2^amp^r}=(\alpha\beta_0)^{2^amp^r}$. Then by Proposition \ref{propostion3.2}, $-\lambda$-constacyclic codes of length $2^am p^{r}$ over $\F_q$ are equivalent to cyclic codes over $\F_q$.
\qed
\begin{cor}
Let $\lambda =\beta_0^{2^amp^r}$, $\alpha$ a primitive $2^a$-th root of unity in $\F_q^*$ and let $C$ be a $\lambda$-constacyclic code of length $2^amp^r$, then:
\[C=\langle (\prod_{k=1}^{2^a}(\prod_{i=0}^{l}f_{i}^{j_i}(\beta_0^{-1}\alpha ^{-k}x))\rangle,\]
where $0\leq j_i\leq p^r$.
\end{cor}
\pf
By Lemma \ref{lem:primitive} if $q \equiv 1 \mod 2^a $, then there exists a primitive $2^a$-th root $\alpha$ of unity in $\F_q^*$.
Thus by Corollary \ref{cor:prod}
\[(x^{2^amp^r}-1)=(x^{2^am}-1)^{p^r}=\prod_{k=1}^{2^a}\prod_{i=0}^{l}f_{i}^{p^r}(\alpha ^{-k}x),\]
then
\[((\beta_0^{-1}x)^{2^amp^r}-1)=((\beta_0^{-1}x)^{2^am}-1)^{p^r}=\prod_{k=1}^{2^a}\prod_{i=0}^{l}f_{i}^{p^r}(\beta_0^{-1}\alpha ^{-k}x),\]
so
\[(x^{2^amp^r}-\lambda)=\lambda((\beta_0^{-1}x)^{2^am}-1)^{p^s}=\prod_{k=1}^{2^a}\prod_{i=0}^{l}f_{i}^{p^r}(\beta_0^{-1}\alpha ^{-k}x).\]
Since any $\lambda$-constacyclic codes of length $2^amp^r$ is generated by a divisor of $(x^{2^amp^r}-\lambda)$
then we have the result.
\qed

\begin{cor}
Let $\lambda =\beta_0^{2^am}$ and let $C$ be a $-\lambda$-constacyclic code of length  $2^amp^r$. If $q \equiv 1 \mod 2^{a+1}$then
\[C=\langle (\prod_{k=1}^{2^{a}}(\prod_{i=0}^{l}f_{i}^{j_i}(\beta_0^{-1} \alpha ^{-2k+1} x)))\rangle\]
where $0\leq j_i\leq p^r$.
\end{cor}
\pf
By Lemma \ref{lem:primitive} if  $q \equiv 1 \mod 2^{a+1} $ there exists a primitive $2^{a+1}$-th root $\alpha$ of unity in $\F_q^*$.
Thus by Corollary~\ref{cor:prod}
\[(x^{2^{a+1}mp^r}-1)=(x^{2^am}-1)^{p^r}(x^{2^am}+1)^{p^r}=\prod_{k=1}^{2^a}\prod_{i=0}^{l}f_{i}^{p^r}(\alpha ^{-2k}x)\prod_{k=1}^{2^a}\prod_{i=0}^{l}f_{i}^{p^r}(\alpha ^{-2k+1}x).\]
Then
\[((\beta_0^{-1}x)^{2^amp^r}+1)=((\beta_0^{-1}x)^{2^am}+1)^{p^r}=\prod_{k=1}^{2^a}(\prod_{i=0}^{l}f_{i}^{p^r}(\beta_0^{-1}\alpha ^{-2k+1}x).\]
Then
\[(x^{2^amp^r}+\lambda)=\lambda((\beta_0^{-1}x)^{2^am}-1)^{p^s}=\prod_{k=1}^{2^a}(\prod_{i=0}^{l}f_{i}^{p^r}(\beta_0^{-1}\alpha ^{-2k+1}x).\]
Since any $-\lambda$-constacyclic codes of length $2^amp^r$ is generated by a divisor of $(x^{2^amp^r}+\lambda)$
 we obtain the result.
\qed

\begin{ex}
Let $n=2\cdot7\cdot5^{3}=1750$, $q=5^{2}$ and $\beta$ be a
primitive element of $\F_{25}^{\star}$. Further, let $\lambda \in \{
\beta ^{2i}, 1\leq i \leq12\}$. Since $(7,24)=1$, $\beta^{2i}$ =
$\beta^{2i(7\cdot7-24\cdot2)}$ = $(\beta^{7})^{2\cdot7i}$.
Then in $\F_{25}$ we have that $x^{7}-1$ = $(x+4)(x^3+\beta x^2+\beta
^{17}x+4)(x^3+\beta^5x^2+\beta^{13}x+4)  =
f_{1}(x)f_{2}(x)f_{3}(x)$. So then $x^{14}-1$ =
$f_{1}(x)f_{2}(x)f_{3}(x)f_{1}(-x)f_{2}(-x)f_{3}(-x)$. Since cyclic
codes of length $2\cdot7\cdot5^3 $ over $\F_{25}$ are ideals of
$\frac{\F_{25}[x]}{x^{1750}-1}$ which is a principal ideal ring. Then 
these codes are generated by
\[
\langle (f_{1}^{s}(x)f_{2}^{j}(x)f_{3}^{k}(x)f_{1}^{l}(-x)f_{2}^{m}(-x)f_{3}^{s}(-x))\rangle, \; s,j,k,l,m,s \in \{0,\ldots, 5^3 \}.
\]
Therefore, $\lambda$-constacyclic  codes of length $2\cdot7\cdot5^3
=1750$ over $\F_{25}$ are ideals of
$\frac{\F_{25}[x]}{x^{1750}-\lambda}$ which is a principal ideal
ring, and these codes are generated by
\[
\begin{array}{l}
\langle (f_{1}^{s}(\beta ^{-11i}x)f_{2}^{j}(\beta ^{-11i}x)f_{3}^{k}(\beta ^{-11i}x)
f_{1}^{l}(-\beta ^{-11i}x)f_{2}^{m}(-\beta ^{-11i}x)f_{3}^{t}(-\beta ^{-11i}x))\rangle,\\
\hspace*{2.3in}\; s,j,k,l,m,t  \in \{0,\ldots ,5^3 \}, 1 \leq i \leq 12.
\end{array}
\]
In $\F_{25}$ we have $7^2=49=-1$, so for $\lambda \in \{ \beta
^{2i}, 1\leq i \leq12\}$, $-\lambda \in \{ (7 \alpha) ^{2i}, 1\leq i
\leq12\}$. Thus $-\lambda$-constacyclic codes of length
$2\cdot7\cdot5^3 =1750$ over $\F_{25}$ are ideals of
$\frac{\F_{25}[x]}{x^{1750}+\lambda}$ which is a principal ideal
ring, and these codes are generated by
\[
\begin{array}{l}
\langle f_{1}^{s}(-7\beta ^{-11i}x)f_{2}^{j}(-7\beta ^{-11i}x)f_{3}^{k}(-7\alpha ^{-11i}x)
f_{1}^{l}(7\beta ^{-11i}x)f_{2}^{m}(7\beta ^{-11i}x)f_{3}^{t}(7 \beta ^{-11i}x)\rangle,\\
\hspace*{2.3in}\; s,j,k,l,m,t \in \{0,\ldots ,5^3\}, 1\leq i \leq 12.
\end{array}
\]
\end{ex}
Hence we obtain Table 1.
\begin{table}
\caption{The Generators Polynomials of $\lambda$-Constacyclic
Codes of Length 1750 over $\F_{25}$.}
\begin{center}
\begin{tabular}{|c|c|c|}
  \hline
  $ \beta^{11i}$& $ \lambda= ( \beta^{11i})^{2\cdot7\cdot5^3 }$ & $f_{1}^{s}(-7\beta ^{-11i}x)f_{2}^{j}(-7\beta ^{-11i}x)f_{3}^{k}(-7\beta ^{-11i}x)$\\&&$f_{1}^{l}(7\beta ^{-11i}x)f_{2}^{m}(7\beta ^{-11i}x)f_{3}^{t}(7 \beta ^{-11i}x)$ \\
  \hline
  $ \beta^{11}$& $(\beta^{11})^{2\cdot7\cdot5^3 }$ & $f_{1}^{s}(-7\beta ^{-11}x)f_{2}^{j}(-7\beta ^{-11}x)f_{3}^{k}(-7\beta ^{-11}x)$\\
  & &$f_{1}^{l}(7\beta ^{-11}x)f_{2}^{m}(7\beta ^{-11}x)f_{3}^{t}(7\beta ^{-11}x)$ \\
  $ \beta^{11\cdot2}$& $(\beta^{11\cdot2})^{2\cdot7\cdot5^3 }$ & $f_{1}^{s}(-7 \beta^{-11\cdot2}x)f_{2}^{j}(-7\beta ^{-11\cdot2}x)
  f_{3}^{k}(-7\beta ^{-11\cdot2}x)$\\&&$ f_{1}^{l}(7 \beta^{-11\cdot2}x)f_{2}^{m}(7\beta ^{-11\cdot2}x)f_{3}^{t}(7 \beta ^{-11\cdot2}x)$ \\
 $ \beta^{11\cdot3}$& $  (\beta^{11\cdot3})^{2\cdot7\cdot5^3 }$ & $f_{1}^{s}(-7 \beta^{-11\cdot3}x)f_{2}^{j}(-7\beta ^{-11\cdot3}x)
  f_{3}^{k}(-7 \beta^{-11\cdot3}x)$\\&&$f_{1}^{l}(7\beta ^{-11\cdot3}x)f_{2}^{m}(7\beta^{-11\cdot3}x)f_{3}^{t}(7\beta ^{-11\cdot3}x)$ \\
  $ \beta^{11\cdot4}$& $  (\beta^{11\cdot4})^{2\cdot7\cdot5^3 }$ & $f_{1}^{s}(-7\beta ^{-11\cdot4}x)f_{2}^{j}(-7\beta ^{-11\cdot4}x)
  f_{3}^{k}(-7\beta ^{-11\cdot4}x)$\\&&$ f_{1}^{l}(7 \beta^{-11\cdot4}x)f_{2}^{m}(7\alpha ^{-11\cdot4}x)f_{3}^{t}(7\beta ^{-11\cdot4}x)$ \\
  $ \alpha^{11\cdot5}$& $  (\beta^{11\cdot5})^{2\cdot7\cdot5^3 }$ & $f_{1}^{s}(-7\beta^{-11\cdot5}x)f_{2}^{j}(-7\beta ^{-11\cdot5}x)
  f_{3}^{k}(-7\beta ^{-11\cdot5}x)$\\&&$ f_{1}^{l}(7\beta ^{-11\cdot5}x)f_{2}^{m}(7\beta ^{-11\cdot5}x)f_{3}^{t}(7\beta ^{-11\cdot5}x)$ \\
  $ \alpha^{11\cdot6}$& $  (\beta^{11\cdot6})^{2\cdot7\cdot5^3 }$ & $f_{1}^{s}(-7\beta ^{-11\cdot6}x)f_{2}^{j}(-7 \beta^{-11\cdot6}x)
  f_{3}^{k}(-7\beta ^{-11\cdot6}x)$\\&&$ f_{1}^{l}(7\beta ^{-11\cdot6}x)f_{2}^{m}(7\beta ^{-11\cdot6}x)f_{3}^{t}(7 \beta^{-11\cdot6}x)$ \\
  $ \alpha^{11\cdot7}$& $  (\beta^{11\cdot7})^{2\cdot7\cdot5^3 }$ & $f_{1}^{s}(-7 \beta^{-11\cdot7}x)f_{2}^{j}(-7\beta ^{-11\cdot7}x)
  f_{3}^{k}(-7\beta ^{-11\cdot7}x)$\\&&$ f_{1}^{l}(7\beta ^{-11\cdot7}x)f_{2}^{m}(7\beta ^{-11\cdot7}x)f_{3}^{t}(7  \beta^{-11\cdot7}x)$ \\
  $ \alpha^{11\cdot8}$& $  (\beta^{11\cdot8})^{2\cdot7\cdot5^3 }$ & $f_{1}^{s}(-7 \beta^{-11\cdot8}x)f_{2}^{j}(-7 \beta^{-11\cdot8}x)
  f_{3}^{k}(-7\beta ^{-11\cdot8}x)$\\&&$ f_{1}^{l}(7\beta ^{-11\cdot8}x)f_{2}^{m}(7\beta ^{-11\cdot8}x)f_{3}^{t}(7\beta ^{-11\cdot8}x)$ \\
  $ \alpha^{11\cdot9}$& $  (\beta^{11\cdot9})^{2\cdot7\cdot5^3 }$ & $f_{1}^{s}(-7 \beta^{-11\cdot9}x)f_{2}^{j}(-7 \beta^{-11\cdot9}x)
 f_{3}^{k}(-7 \beta^{-11\cdot9}x)$\\&&$ f_{1}^{l}(7\beta ^{-11\cdot9}x)f_{2}^{m}(7 \beta^{-11\cdot9}x)f_{3}^{t}(7  \beta^{-11\cdot9}x)$ \\
  $ \alpha^{11\cdot10}$& $  (\beta^{11\cdot10})^{2\cdot7\cdot5^3 }$ & $f_{1}^{s}(-7 \beta^{-11\cdot10}x)f_{2}^{j}(-7 \beta^{-11\cdot10}x)
  f_{3}^{k}(-7 \beta^{-11\cdot10}x)$\\&&$ f_{1}^{l}(7\beta ^{-11\cdot10}x)f_{2}^{m}(7\beta ^{-11\cdot10}x)f_{3}^{t}(7\beta ^{-11\cdot10}x)$ \\
  $ \alpha^{11\cdot11}$& $  (\beta^{11\cdot11})^{2\cdot7\cdot5^3 }$ & $f_{1}^{s}(-7\beta ^{-11\cdot11}x)f_{2}^{j}(-7 \beta^{-11\cdot11}x)
  f_{3}^{k}(-7\beta^{-11\cdot11}x)$\\&&$ f_{1}^{l}(7\beta ^{-11\cdot11}x)f_{2}^{m}(7 \beta^{-11\cdot11}x)f_{3}^{t}(7  \beta^{-11\cdot11}x)$ \\
  $ \alpha^{11\cdot12}$& $  (\beta^{11\cdot12})^{2\cdot7\cdot5^3 }$ & $f_{1}^{s}(-7\beta ^{-11\cdot12}x)f_{2}^{j}(-7 \beta^{-11\cdot12}x)
  f_{3}^{k}(-7\beta ^{-11\cdot12}x)$\\&&$ f_{1}^{l}(7\beta ^{-11\cdot12}x)f_{2}^{m}(7 \beta^{-11\cdot12}x)f_{3}^{t}(7 \beta^{-11\cdot12}x)$ \\

  \hline
\end{tabular}
\end{center}
\end{table}
\section{Self-Dual Negacyclic Codes of Length $2^amp^r$ over $\mathbb{F}_{p^s}$}

Let $p$ be an odd prime number and $n=mp^r$,
with $m$ an integer (odd or even) such that $(m,p)=1$. This section
provides conditions on the existence of self-dual negacyclic codes
of length $n=2^amp^r$ over $\mathbb{F}_{p^s}$. It is well known that negacyclic
codes over $\mathbb{F}_{p^s}$ are principal ideals generated by the factors
of $x^{mp^r}+1$. Since $\mathbb{F}_{p^s}$ has characteristic $p$, and negacyclic codes are
a particular case of constacyclic codes, so by results of Section \ref{section:3}, the
polynomial $x^{mp^r}+1$ can be factored as
\begin{equation}
\label{eq:factor2} x^{mp^r}+1=(x^{m}+1)^{p^r}.
\end{equation}
The polynomial $x^{m}+1$ is a monic square free polynomial. Hence
from~\cite[Proposition 2.7]{permounth} it factors uniquely as a
product of pairwise coprime monic irreducible polynomials $f_0(x),
\ldots, f_l(x)$. Thus from~(\ref{eq:factor2}) we obtain the
following factorization of $x^{mp^r}+1$
\begin{equation}
\label{eq:factor3} x^{mp^r}+1={f_0(x)}^{p^r}\ldots {f_l(x)}^{p^r}.
\end{equation}
A negacyclic code of length $n=mp^r$ over $\mathbb{F}_{p^s}$ is then
generated by a polynomial of the form
\begin{equation}
\label{eq:gen}
 A(x)=\prod {f_i}^{k_i},
 \end{equation}
where $f_i(x), 0 \leq i \le l$, are the polynomials given in
(\ref{eq:factor3}) and $0 \le k_i \le p^r$.

For a polynomial $f(x)=a_0+a_1 x\ldots+ a_rx^r$, with $a_0 \neq 0$
and degree $r$ (hence $a_r\neq 0$), the reciprocal of $f$ is the
polynomial denoted by $f^*$ and defined as
\begin{equation}
\label{eq:5} f^*(x)=x^rf(x^{-1})=a_r+a_{r-1}x+\ldots + a_0x^r.
\end{equation}
If a polynomial $f(x)$ is equal to its reciprocal, then $f(x)$ is called
self-reciprocal. We can easily verify the following equalities
\begin{equation}
\label{eq:prop}
 (f(x)^*)^*=f (x)\text{ and } (fg(x))^*=f(x)^*g(x)^*.
\end{equation}

It is well known (see \cite[Proposition 2.4]{dinh1}, that the dual of the negacyclic code generated
by $A(x)$ is the negacyclic code generated by $B^*(x)$ where
\begin{equation}
\label{eq:dual} B(x)=\frac{x^n+1}{A(x)}.
\end{equation}
Hence we have the following lemma.
\begin{lem}
\label{lem:dual} A negacyclic code $C$ of length $n$ generated by a
polynomial $A(x)$ is self-dual if and only if
\[
A(x)=B^*(x).
\]
\end{lem}
Denote the factors $f_i$ in the factorization of $x^m+1$ which are
self-reciprocal by $g_1, \ldots g_s$, and the remaining $f_j$
grouped in pairs by $h_1,h_1^*,\ldots, h_t,h_t^*$. Hence $l=s+2t$
 and the factorization given in~(\ref{eq:factor3}) becomes
\begin{equation}
\begin{array}{ccl}
\label{eq:ling} x^n+1&=&(x^m+1)^{p^r}=g^{p^r}_1(x)\ldots
g^{p^r}_s(x)\\
&&\times h^{p^r}_1(x)h^{* p^r}_1(x)\ldots h^{p^r}_t(x)h^{*
p^r}_t(x).
\end{array}
\end{equation}
In the following  we give the structure of negacyclic codes over
$\mathbb{F}_{p^s}$ of length $2^amp^r$, $a\geq1$.
We begin with this useful lemma.
\begin{lem}
\label{lem:equiv} Let $q=p^s$ be an odd prime power such that $q\equiv
1 \mod 2^{a+1}$. Then there is a ring isomorphism between the ring $\frac{
\mathbb{F}_{q}[x]}{x^{2^amp^r}-1}$ and the ring $\frac{\mathbb{F}_{q}[x]}{x^{2^amp^r}+1}$
\end{lem}
\pf
If  $q\equiv 1 \mod 2^{a+1}$ then by Lemma~\ref{lem:primitive} there exists a primitive $2^{a+1}$-th root $\alpha$ of unity in $\F_q^*$.
So $-1=(-1)^{mp^r}=(\alpha^{2^a})^{mp^r}$ and then by Proposition \ref{propostion3.2}, negacyclic codes of length $2^am p^{r}$ over $\F_q$ are equivalent to cyclic codes of length $2^am p^{r}$ over $\F_q$.
\qed

\begin{cor}
\label{cor:product} Let  $q=p^s$ be an odd prime power such that $q\equiv
1 \mod 2^{a+1}$ and $n=2^amp^r$ with $m$ an odd integer such that $(m,p)=1$. Then a negacyclic
code of length $n$ over $\mathbb{F}_{p^s}$ is a principal ideal of
$\mathbb{F}_{p^s}[x]/\langle x^n+1 \rangle$ generated by a polynomial of the
following form
$\prod_{k=1}^{2^a}(\prod_{i=0}^{l}f_{i}^{j_i}(\alpha ^{-2k+1}x))$
where $0\leq j_i\leq p^r$ and  $f_i(x)$ are monic irreducible factors of $x^m-1$.
\end{cor}

\pf  It suffices to find the factors of $x^{2mp^r}+1$.
Since  $q\equiv 1 \mod 2^{a+1}$ and from~Lemma~\ref{lem:primitive}, there exist $\alpha\in \F_q^*$,
a primitive $2^{a+1}$-th root of unity.So $x^{2mp^r}+1$ can be decomposed as
$(x^{2m}+1)^{p^r}=(\prod_{k=1}^{2^a}\prod_{i=0}^{l}f_{i}(\alpha ^{-2k+1}x))^{p^r}$. The
result then follows from the isomorphisms given in~Lemma~\ref{lem:equiv}.

We recall the most important result of \cite{GG}.
\begin{thm}(  \cite[Theorem 2.2]{GG})
\label{th:exist} There exists a self-dual negacyclic code of length
$mp^r$ over $\mathbb{F}_{p^s}$ if and only if there is no $g_i$
(self-reciprocal polynomial) in the factorization of $x^{mp^r}+1$
given in~(\ref{eq:ling}). Furthermore, a self-dual negacyclic code
$C$ is generated by a polynomial of the following form
\begin{equation}
h^{b_1}_1(x)h^{* p^r- b_1}_1(x)\ldots h^{b_t}_t(x)h^{*
p^r-b_t}_t(x).
\end{equation}
\end{thm}
\qed

In the following we generalize \cite[Theorem 3.7]{GG} for the length $2^amp^r$.
But before that we need the following lemmas.
\begin{lem}(\cite[Lemma 3.5]{GG})
\label{lem:rev1} Let $m$ be an odd integer and $Cl_m(i)$ the $p^s$
cyclotomic class of $i$ modulo $m$. The polynomial $f_i(x)$ is the
minimal polynomial associated with $Cl_m(i)$, hence we have
$Cl_m(i)=Cl_m(-i)$ if and only if $f_i(x)=f_i^*(x)$.
\end{lem}
\begin{lem}( \cite[Lemma 3.6]{GG})
\label{lem:rev2} Let $m$ be an odd integer and $p$ a prime number.
Then $\ord_m(p^s)$ is even if and only if there exists a cyclotomic
class $Cl_m(i)$ which satisfies $Cl_m(i)=Cl_m(-i)$.
\end{lem}

\begin{thm}
\label{th:self} Let $q=p^s$ be an odd prime power such that $q\equiv
1 \mod 2^{a+1}$, and $n=2^amp^r$ be an  integer with $(m,p)=1$ and $a>1$. Then there exists
a negacyclic self-dual code of length $2^amp^r$ over $\mathbb{F}_{p^s}$ if and
only if $\ord_m(q)$ is odd.
\end{thm}
\pf  Under the hypothesis on $q$,  and $m$ we have from
Corollary  ~\ref{cor:product} that the polynomial $x^{2mp^r}+1= \prod_{k=1}^{2^a}\prod_{i=0}^{l}f_{i}^{j_i}(\alpha ^{-2k+1}x))$
, where $f_i(x)$ are the monic irreducible factors of $x^m-1$ in
$\mathbb{F}_{p^s}$. By Lemma~\ref{lem:rev2}, $\ord_m(p^s)$ is odd if and
only if there is no cyclotomic class such that $Cl_m(i)=Cl_m(-i)$.
From Lemma~\ref{lem:rev1}, this is equivalent to saying that there
are no irreducible nontrivial factors of $x^m-1$ such that $f_i(x)=f_i^*(x)$.
From corollary ~Corollary ~\ref{cor:product}, we obtain
that $f_i(x)\ne f_i^*(x)$ for all $i\neq 0$ ($f_0(x)=(x-1)$)is true if and only if
$f_{i}( \alpha ^{k}x)\neq f^*_i(\alpha ^{k} x)$ are true for all $1\leq k \leq 2^{a+1}$.
Then from~Theorem \ref{th:exist} self-dual negacyclic codes exist.
\qed

\begin{ex}
A self-dual negacyclic code of length $70$ over $\mathbb{F}_5$ does not
exist. There is no self-dual negacyclic code of length 30 over
$\mathbb{F}_9$, but there is a self-dual code over $\mathbb{F}_9$ of length $126$.
\end{ex}

\end{document}